\documentclass[lettersize,journal,compsoc]{IEEEtran}
\usepackage{amsmath,amsfonts}
\usepackage{algorithmic}
\usepackage{algorithm}
\usepackage{array}
\usepackage[caption=false,font=normalsize,labelfont=sf,textfont=sf]{subfig}
\usepackage{textcomp}
\usepackage{stfloats}
\usepackage{url}
\usepackage{verbatim}
\usepackage{graphicx}
\usepackage{hyperref}

\pdfoutput=1

\ifCLASSOPTIONcompsoc
  \usepackage[nocompress]{cite}
\else
  \usepackage{cite}
\fi

\AtBeginDocument{%
  \providecommand\BibTeX{{%
    \normalfont B\kern-0.5em{\scshape i\kern-0.25em b}\kern-0.8em\TeX}}}

\usepackage{etoolbox}
\makeatletter
\patchcmd{\@makecaption}
  {\scshape}
  {}
  {}
  {}
\makeatletter
\patchcmd{\@makecaption}
  {\\}
  {.\ }
  {}
  {}
\makeatother

\usepackage[table,xcdraw]{xcolor}
\usepackage{multirow}
\usepackage{cite}
\usepackage{amsmath,amssymb,amsfonts}
\usepackage{algorithmic}
\usepackage{graphicx}
\usepackage{svg}

\usepackage{balance}

\usepackage{textcomp}
\usepackage{xcolor}
\def\BibTeX{{\rm B\kern-.05em{\sc i\kern-.025em b}\kern-.08em
    T\kern-.1667em\lower.7ex\hbox{E}\kern-.125emX}}
    
\AtBeginDocument{%
  \providecommand\BibTeX{{%
    Bib\TeX}}}

\def\code#1{\texttt{#1}}

\usepackage{enumitem}
\usepackage{multirow}
\usepackage{hhline}
\usepackage{graphicx}
\usepackage{svg}
\usepackage{colortbl}
\usepackage{balance}

\newcommand{\contradict}{\textcolor{red}{$\times$} }
\newcommand{\unsupported}{$\bigcirc$ }

\newcommand{\revision}[1]{\textcolor{blue}{#1}}

\definecolor{blue_}{HTML}{B8EBFF}
\definecolor{pink_}{HTML}{FFCCCB}
\definecolor{umber}{HTML}{CC0000}
\definecolor{orange}{HTML}{CC5500}
\definecolor{redd}{HTML}{F26035} 

\definecolor{asparagus}{rgb}{0.53, 0.66, 0.42}

\newcommand{\positive}[1]{\colorbox{blue_}{#1}}
\newcommand{\negative}[1]{\colorbox{pink_}{#1}}
\newcommand{\graybg}[1]{\colorbox{gray!25}{#1}}

\usepackage{tikz}
\newcommand{\greencheck}{}%
\DeclareRobustCommand{\greencheck}{%
  \tikz\fill[scale=0.4, color=teal]
  (0,.35) -- (.25,0) -- (1,.7) -- (.25,.15) -- cycle;%
}

\usepackage{textcomp}
\usepackage{xcolor}
\def\BibTeX{{\rm B\kern-.05em{\sc i\kern-.025em b}\kern-.08em
    T\kern-.1667em\lower.7ex\hbox{E}\kern-.125emX}}
    
\newenvironment{boxedtext}
    {
    
    \begin{center}

    \begin{tabular}{|p{0.96\linewidth}|}
    \hline
    }
    { 
    \\ \hline
    \end{tabular} 
    
    \end{center}
       }

\definecolor{purple_col}{HTML}{CBCEFB}
\definecolor{orange_col}{HTML}{FED8B1}

\def\code#1{\texttt{#1}}

\newcolumntype{L}[1]{>{\raggedright\let\newline\\\arraybackslash\hspace{0pt}}m{#1}}
\newcolumntype{C}[1]{>{\centering\let\newline\\\arraybackslash\hspace{0pt}}m{#1}}
\newcolumntype{R}[1]{>{\raggedleft\let\newline\\\arraybackslash\hspace{0pt}}m{#1}}

\begin{document}

\title{From First Patch to Long-Term Contributor: Evaluating Onboarding Recommendations for OSS Newcomers }

\author{Asif~Kamal~Turzo,
        Sayma~Sultana,       
        Amiangshu Bosu,~\IEEEmembership{Member,~IEEE}
\IEEEcompsocitemizethanks{\IEEEcompsocthanksitem A. Turzo, S. Sultana, and A. Bosu are with the Department
of Computer Science, Wayne State University, Detroit,
MI, 48202.\protect\\
E-mail: asifkamal@wayne.edu, sayma@wayne.edu, and abosu@wayne.edu}
\thanks{Manuscript received July 4, 2024; Revised January 5, 2025 }}


\maketitle

\begin{abstract}
Attracting and retaining a steady stream of new contributors is crucial to ensuring the long-term survival of open-source software (OSS) projects. However, there are two key research gaps regarding recommendations for onboarding new contributors to OSS projects. 
First, most of the existing recommendations are based on a limited number of projects, which raises concerns about their generalizability. If a recommendation yields conflicting results in a different context, it could hinder a newcomer's onboarding process rather than help them.
Second, it's unclear whether these recommendations also apply to experienced contributors. If certain recommendations are specific to newcomers, continuing to follow them after their initial contributions are accepted could hinder their chances of becoming long-term contributors.
To address these gaps, we conducted a two-stage mixed-method study. In the first stage, we conducted a Systematic Literature Review (SLR) and identified 15 task-related actionable recommendations that newcomers to OSS projects can follow to improve their odds of successful onboarding. In the second stage, we conduct a large-scale empirical study of five Gerrit-based projects and 1,155 OSS projects from GitHub to assess whether those recommendations assist newcomers' successful onboarding. Our results suggest that four recommendations positively correlate with newcomers' first patch acceptance in most contexts. Four recommendations are context-dependent, and four indicate significant negative associations for most projects. Our results also found three newcomer-specific recommendations, which OSS joiners should abandon at non-newcomer status to increase their odds of becoming long-term contributors.

\end{abstract}

\begin{IEEEkeywords}
newcomer, onboarding, software development, open source software
\end{IEEEkeywords}

\IEEEoverridecommandlockouts

\section{Introduction}
\label{sec:intro}

To thrive, Open Source Software (OSS) projects must attract and retain new contributors \cite{crowston2003defining}. Well-known OSS projects like Apache, Linux, and Android attract enthusiastic candidates who aim to become long-term contributors~\cite{von2003community,lakhani2003hackers,alexander2002working}. However, due to the challenging nature of the OSS onboarding process~\cite{steinmacher2015social,steinmacher2014barriers}, most of these candidates fail to become long-term contributors~\cite{von2003community,ducheneaut2005socialization}.

For an aspiring OSS candidate, having the first patch accepted is a significant milestone towards this goal. The sense of accomplishment and recognition of the first accepted patch motivates further contributions~\cite{ducheneaut2005socialization,carillo2017makes,steinmacher2015systematic}. Several studies~\cite{tan2020first,choi2010socialization,steinmacher2018let} have focused on prospective OSS contributors who have not yet had changes accepted ~\cite{steinmacher2015social,subramanian2020analyzing}. According to Steinmacher \textit{et al.}, a person is considered a `newcomer' to an OSS project until he/she gets his/her first code review/pull request accepted to that project \cite{steinmacher2015social}.  Studies have proposed recommendations to increase newcomers' chances of successful onboarding. 

However, there are two uncertainties regarding the recommendations based on current research. First, due to the diverse nature of OSS, it's uncertain how well these recommendations will apply to other OSS projects. Let's consider the scenario where a recommendation provides contradicting results in a different context. Following this recommendation in an unhelpful context may hinder a newcomer's onboarding instead of helping.
Second, it remains unclear whether the same recommendations also apply to non-newcomers. If some recommendations are specific to newcomers, contributors who continue to follow them after their first changes are accepted could harm their chances of becoming long-term contributors.
Therefore, this study addresses these two research gaps by evaluating newcomers' onboarding recommendations i) across a diverse group of OSS projects to assess whether they are context-dependent and ii) on non-newcomers to assess whether they are newcomer-specific.

\begin{figure}
	\centering  \includegraphics[width=.95\linewidth]{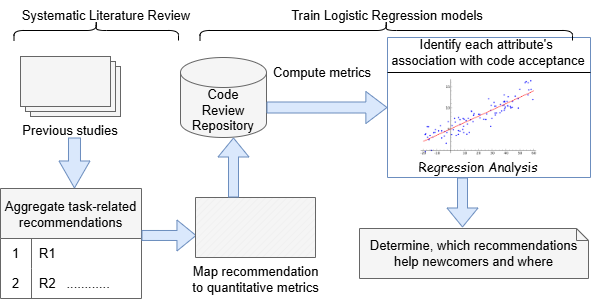}
	\caption{An overview of our two-stage mixed method research design}
	\label{fig:approach_overview}	
\end{figure}

Figure ~\ref{fig:approach_overview} provides an overview of our two-stage mixed-method empirical study towards these goals. This study specifically focuses on task-related recommendations since they apply to all newcomers and are easily implementable. On the contrary, recommendations such as having expertise in a project's language or framework may take time, and an OSS newcomer may not yet be motivated to spend time-consuming efforts gaining those due to uncertainties of onboarding.
In the first stage, we conducted a systematic literature review (SLR) to curate a list of task-related recommendations based on existing research.  In the second stage, we identify a quantitative metric to measure a newcomer's compliance with each recommendation. Finally,  we conducted a large-scale empirical study of five OSS projects using Gerrit-based code reviews (CR) and 1,155 projects using GitHub's pull requests (PR) to evaluate the association between a CR/PR's compliance with various recommendations and its likelihood of acceptance.
We investigate the following three research questions.

\vspace{2pt}
\noindent \textbf{(RQ1) What are the task-related recommendations for prospective newcomers of 
OSS projects?}

\noindent \underline{Motivation:}  Newcomer onboarding is essential for integrating new ideas and ensuring the sustainability of OSS projects. Due to its importance, prior studies have identified several recommendations to assist the onboarding of OSS newcomers \cite{subramanian2020analyzing,balali2020recommending,rehman2022newcomer,li2021you}. 
 While our primary motivation behind this question is to identify the recommendations and validate them in our RQ2 empirically, answering this question will also aggregate these recommendations into a common list that a newcomer can examine. 
 
 \vspace{2pt}
  \noindent \underline{Method:} Following the guidelines of Kitchenham \textit{et} al.~\cite{kitchenham2015evidence}, we conducted a Systematic Literature Review (SLR) that identified 16 primary studies. 
  
 \vspace{2pt} 
 \noindent \underline{Key results:} We have identified 15 task-related recommendations for prospective newcomers to OSS projects, which include nine recommendations on task selection, four on task preparation, and two regarding task completion.
 
\vspace{2pt}
 \noindent \textbf{(RQ2) How do compliance with OSS newcomers' onboarding recommendations affect the odds of acceptance of their first patches?}

\vspace{2pt}
 \noindent \underline{Motivation:} Although prior studies suggest guidelines for newcomers to assist their onboarding, there is a lack of large-scale evaluation of their efficacy under various OSS contexts. Moreover, many recommendations are based on group opinions (e.g., interviews or surveys) and have not been quantitatively evaluated.
 Answering RQ2 will help newcomers and OSS mentors identify context-specific applicable recommendations to increase the odds of successful onboarding.

\vspace{2pt}
 \noindent \underline{Method:}  We defined a quantitative metric for each recommendation to measure a newcomer's compliance against it.   We constructed logistic regression models following the suggestions of Harrell Jr.~\cite{harrell2015regression}, where code acceptance is the dependent and 15 recommendation compliance metrics are factors.

\vspace{2pt}
 \noindent \underline{Key results:} 
 The results of our regression models suggest that four identified recommendations have significant positive associations with newcomers' first patch acceptance in most contexts. Four recommendations are context-dependent, four indicate significant negative associations in most projects, and the remaining ones are either inconclusive or could not be evaluated due to multicollinearity (defined in Section~\ref{model-training}).

 \vspace{2pt}
 \noindent \textbf{(RQ3) Do OSS newcomers' onboarding recommendations also apply to non-newcomers? }

\vspace{2pt}
 \noindent \underline{Motivation:} 
Identifying recommendations that do not have any significant association or decrease the odds of acceptance for non-newcomers will help an OSS contributor with few accepted patches choose a strategy to improve their chances of becoming long-term contributors.

\vspace{2pt}
 \noindent \underline{Method:}  
 Using the same set of metrics and regression modeling strategy as RQ2, we evaluate compliance of those recommendations against the odds of a non-newcomer's patch acceptance. 

\vspace{2pt}
 \noindent \underline{Key results:} 
Our results found three newcomer-specific recommendations contradicting their associations between newcomers and non-newcomers. 
Therefore, OSS candidates should avoid using these three strategies once their first patches are accepted to enhance their chances of becoming long-term contributors.

\vspace{2pt}
\noindent \textbf{Contributions:}
 Primary contributions of this study include:
\begin{itemize}
  \item A Systematic Literature Review (SLR) to aggregate the task-related recommendations from the existing literature to assist newcomers' successful onboarding.
  \item A list of quantitative metrics to evaluate compliance with newcomers' task-related recommendations.
  \item An empirical assessment of whether newcomers' task-related recommendations increase their odds of successful onboarding.
  \item Recommendations to newcomers on their post-onboarding task selection to assist long-term success with a project.
  \item We have released our dataset and scripts on \href{https://github.com/WSU-SEAL/newcomer-onboarding}{GitHub}\footnote{\url{https://github.com/WSU-SEAL/newcomer-onboarding}}.
\end{itemize}

\vspace{2pt}
\noindent \textbf{Organization:}
The remainder of the paper is organized as follows. Section \ref{sec:background} briefly overviews the related works.
Section \ref{sec:slr} details the methodology and results of our first stage (i.e., SLR).
Section \ref{sec:stage2} details the methodology and results of empirical evaluation to investigate RQ2 and RQ3.
Section \ref{sec:discussion} discusses key implications based on our results.   Section \ref{sec:validity} addresses threats to validity for our findings.
Finally, Section \ref{sec:conclusion} concludes this paper.

\section{Related Works}
\label{sec:background}

 \vspace{4pt} \noindent \textbf{Challenges faced by newcomers: } A newcomer's journey to becoming a long-term contributor is full of challenges~\cite{steinmacher2015social, hannebauer2014exploratory, guizani2021long, pinto2016more, steinmacher2014hard,steinmacher2014support,steinmacher2014attracting}. Steinmacher \textit{et} al. identified  58 social barriers newcomers face for contributing their first change \cite{steinmacher2015social}. Besides social ones, newcomers also encounter technical challenges in contributing to large, matured OSS projects \cite{guizani2021long}.  Prior studies also studied barriers faced by one-time contributors in OSS projects, who had exactly one patch accepted \cite{lee2017understanding},\cite{lee2017one}, quasi-contributors (newcomers without any accepted CR/PR)\cite{steinmacher2018almost}, and casual contributors who contribute occasionally~\cite{pinto2016more}.  Another couple of studies focused on drive-by commits~\cite{pham2013building} and episodic volunteering\cite{barcomb2020managing}.
 These studies conclude that the primary challenges encountered by prospective OSS joiners include lack of support materials~\cite{lee2017understanding,steinmacher2015systematic,steinmacher2014barriers,steinmacher2021being}, a steep learning curve~\cite{guizani2021long,hannebauer2014exploratory, steinmacher2014barriers}, unfriendly project maintainers~\cite{steinmacher2014barriers,lee2017understanding,balali2018newcomers}, and lack of timely feedback~\cite{bosu2014impact,lee2017understanding,steinmacher2014barriers}.

\vspace{4pt} \noindent \textbf{Recommendations to assist newcomers' onboarding: }
 Several existing studies have identified recommendations to assist newcomers' onboarding process.
 These studies have focused on various aspects that a newcomer can follow.
 Studies have focused on
\textbf{Task selection}, i.e., which task a newcomer should start with~\cite{subramanian2020analyzing,guizani2021long,santos2021can,santos2022choose,horiguchi2021onboarding}. Some studies have also focused on \textbf{Task preparation}, i.e., how to prepare the changeset to facilitate easy reviewing from maintainers~\cite{xiao2022recommending,tan2020first}.  \textbf{Task completion}-related suggestions focus on communicating and responding to reviewers' suggestions after a change is submitted for reviews~\cite{balali2018newcomers,steinmacher2016overcoming,guizani2021long}.  \textbf{Socialization and communication}-related guidelines focus on etiquette and strategies to connect with existing members  ~\cite{steinmacher2018let,steinmacher2016overcoming,steinmacher2021being,trinkenreich2020hidden}.
 Besides newcomers themselves, project maintainers also play important roles in fostering a supportive environment for newcomers \cite{zhou2012make}.
  Recommendations for maintainers include  mentoring~\cite{balali2020recommending}, creating a documentation outlining step-by-step onboarding process~\cite{sholler2019ten,steinmacher2016overcoming,fronchetti2023contributing}, ensuring that newcomers feel welcome~\cite{sholler2019ten}, having and enforcing code of conduct~\cite{sholler2019ten,li2021code,steinmacher2016overcoming},
  implementing automated task recommendation system~\cite{xiao2022recommending}, using specialized tags for tasks that newcomers can work on~\cite{huang2021characterizing},   using a community health dashboard to track newcomers' progress~\cite{guizani2022attracting}, and creating a portal to help newcomers overcome common entry barriers \cite{steinmacher2016overcoming}.

\vspace{4pt} \noindent \textbf{Automated tools for assisting newcomers:}
Proper mentoring throughout the onboarding journey is essential to ensure newcomers' future success \cite{fagerholm2014role,fagerholm2013onboarding,fagerholm2014onboarding}. To assist the newcomers in this journey, researchers have proposed an automated model to identify potential mentors~\cite{canfora2012going}, conversational bots that can act as a mentor \cite{dominic2020conversational, dominic2020onboarding} and answer newcomers' questions~\cite{malheiros2012source}. 
Another direction focused on building models to automatically identify suitable tasks or bugs that newcomers can start with \cite{hegfi,xiao2022recommending,wang2011bug,kallis2019ticket,izadi2022predicting,santos2021can,huang2021characterizing}. Since many OSS projects have steep learning curves~\cite{steinmacher2014barriers}, researchers have proposed tools to help newcomers gain knowledge of project goals and design by building project-specific knowledge-based project memory \cite{cubranic2005hipikat,berlin1993did}, a protocol for exchanging information via integrated development environment \cite{box2000simple}, and reverse-engineered visualization tools~\cite{park2009beyond}.

\vspace{4pt} \noindent \textbf{Novelty of this study:}
Our study differs from prior works based on three novelties. First, we quantitatively assess all recommendations on a wide range of OSS projects. Such assessments are crucial since many recommendations are based on qualitative investigations.  Second, we investigate if context dependence applies to any of these recommendations. Third, we aim to assess if those recommendations also apply to non-newcomers to assist the long-term task strategies of OSS candidates. 

\vspace{4pt} \noindent \textbf{Closely related work:}
Tan \textit{et al.}'s study to investigate the characteristics of Good First Issues (GFI) on GitHub~\cite{tan2020first} is the closest one to ours. They examined correlations between a GFI effectiveness (i.e., indeed solved by a newcomer) and its resolution process,  contributors' characteristics, textual characteristics, and labelers' reputations.
As Tan \textit{et} al.'s characterization of effective GFIs overlaps with this study's Task Selection aspect, five of their textual metrics also belong among the 15 selected in our study. 
However,  this study differs from Tan \textit{et} al.~\cite{tan2020first} in the following three key areas. First, this study aims to help newcomers select, prepare, and complete tasks. On the other hand, Tan \textit{et} al.~\cite{tan2020first} aims to help maintainers label more effective GFIs. Second, we examine any change newcomers worked on, including GFIs and non-GFIs. Finally, we also empirically evaluate how newcomers' and non-newcomers' task preparation and completion practices influence code acceptance.

\section{Stage 1: Identification of task-related recommendations for newcomers}
\label{sec:slr}
The following subsections detail our systematic literature review (SLR) methodology to identify a set of task-related onboarding recommendations and 
 SLR results.

\subsection{SLR Methodology}
We conducted an SLR  of existing studies on OSS newcomers' onboarding to identify task-related recommendations that a prospective newcomer can follow. We found two existing SLRs focusing on newcomers' barriers~\cite{steinmacher2015systematic,steinmacher2014barriers}. Since the previous SLRs were not conducted in the context of newcomers' task-related recommendations and several recent studies have recommended new strategies to assist newcomers' onboarding \cite{xiao2022recommending,tan2020first,subramanian2020analyzing,steinmacher2018let}, we decided to conduct a new  SLR.
Our SLR specifically focuses on finding recommendations regarding newcomers'  task selection and work practices, as those are crucial factors behind a newcomer's successful onboarding~\cite{steinmacher2018let}. We do not focus on recommendations that are targeted toward project maintainers. 
We designed our SLR based on the guidelines suggested by Kitchenham \textit{et} al.  \cite{kitchenham2015evidence}, and it has been the gold standard for conducting SLR in the SE domain. Table~\ref{table:number_of_papers} provides the number of papers surviving each SLR step. The steps are as follows.

\begin{table}
	\centering
    \caption{Overview of our SLR steps}    
	\label{table:number_of_papers}
	\resizebox{\linewidth}{!}{
\begin{tabular}{|p{6.5cm}|p{1.2cm}|}
\hline
\multicolumn{1}{|c|}{\textbf{Paper selection steps}}   & \multicolumn{1}{c|}{\textbf{\# of papers}}  \\ \hline
Searching Google Scholar, IEEE, ACM, and  Scopus databases  &  {605}  \\ \hline
After removing duplicates      &   {498}   \\ \hline
Study selection based on title and abstract   &  {68}  \\ \hline
Study selection by a quick reading of the publications     &   {32}  \\ \hline
Inclusion of additional  studies based on Snowballing   &   {40}  \\ \hline
After adding studies based on expert suggestions    & {45}  \\ \hline
Studies surviving all eight exclusion criteria  & {16} \\ \hline

\end{tabular}
}%
\end{table}

 \begin{table} 
     \centering
      \caption{Search keywords for our SLR}
     \resizebox{\linewidth}{!}{
         \begin{tabular}{|p{1.5cm}|p{6cm}|}
         \hline
         \textbf{Category} & \textbf{Keywords} \\ \hline
             Target person &  newcomer OR joiner OR newbie OR beginner OR novice OR ``new committer'' OR ``new contributor'' OR ``potential participant'' \\ \hline
             Goal &   onboarding OR ``good first issue'' OR ``joining process'' OR recommendation OR recommending  OR contribute OR  contributing OR entry OR task OR ``pull request''\\ \hline
             Context & ``open source'' OR ``OSS'' OR ``free software'' OR ``OSS'' OR FLOSS \\ \hline
         \end{tabular}}    
     \label{tab:search_keywords}
 \end{table}


\vspace{-4pt}
\subsubsection*{Step 1: Define inclusion-exclusion criteria}
Our primary goal of performing the SLR is to \textit{identify recommendations regarding newcomers'  task selection and work practices}. For this goal, we defined four inclusion criteria (IC) and eight exclusion criteria (EC). 
Our inclusion criteria are:
\begin{itemize}
    \item \textbf{IC1:} The publication provides recommendation(s) to newcomers of OSS projects regarding task selection.
    \item \textbf{IC2:} The publication provides recommendation(s) to newcomers to OSS projects regarding work practice such as task preparation, task presentation, and communication.
    \item \textbf{IC3:} The research work has been published in peer-reviewed workshops, conferences, or journals.
    \item \textbf{IC4:} The publication is written in English.

 \end{itemize}
    
\noindent We defined the following eight exclusion criteria (EC). 
 \begin{itemize}
     
    \item \textbf{EC1:} The study is an MS or Ph.D. thesis that is not published as a paper.
    \item \textbf{EC2:} Any publication that is not peer-reviewed.
    \item \textbf{EC3:} Any publication we can not fully access online.
    \item \textbf{EC4:} Any study that discusses only machine learning or information retrieval approaches to automatically recommend tasks for easier onboarding.
    \item \textbf{EC5:} Any study that only discusses recommendations for newcomer onboarding based on results from other books, published studies, or articles.
    
    \item \textbf{EC6:} Any study that provides recommendations only for the OSS project maintainers.
    \item \textbf{EC7:} Any study unrelated to OSS.
    \item \textbf{EC8:} None of the included recommendations are related to task selection or work practice.
\end{itemize}

\subsubsection*{Step 2: Search digital libraries}

We adopted a hybrid approach for performing SLR, which involves searching in Google Scholar, Scopus, IEEE, and ACM. Following the suggestions of prior studies \cite{steinmacher2015social,steinmacher2015systematic} and considering our final goal, we defined our search strings using three categories of keywords, which include i) target audience, ii) goal, and iii) context. 
We have used the `advanced search' option of the above-mentioned search engines.
 Table ~\ref{tab:search_keywords} presents the three categories and lists the keywords belonging to each. While the search queries vary among the engines, the queries were formulated in such a way that each result must include at least one keyword from each of the three categories. A simplified search string looks like the following.

 \begin{boxedtext}
(``open source'' OR OSS OR ``free software'' OR OSS OR FLOSS) AND (newcomer OR joiner OR newbie OR beginner OR novice OR ``new committer'' OR ``new contributor'' OR ``potential participant'') AND ( onboarding OR ``good first issue'' OR ``joining process'' OR recommend OR contribute OR entry OR task OR ``pull request'')
\end{boxedtext}

We noticed that the number of entries returned by Google Scholar was large. As most of the results are irrelevant, inspecting all entries would have been extremely time-consuming. Therefore, we went through the results pages containing ten entries individually. If the title of a paper seemed relevant, we stored the title and list of authors in a spreadsheet for further inspection. We stopped inspecting further results if five consecutive pages returned no relevant titles, as Google Scholar ranks search results by the relevance of search query \cite{martin2018google}. 
We conducted two database searches. Our first search was conducted during the last week of August 2022. In the first search, we selected papers based on study titles from the results. However, this protocol is not fully compliant with Kitchenham \textit{et} al.'s guidelines~\cite{kitchenham2015evidence}, which recommend inspecting both titles and abstracts before excluding any search result. 
Therefore, we conducted a fully compliant second search during the second week of October 2024. Moreover, this second search also helped us include papers published since our first search.
Our database search found 605 papers, of which 349 were found in the ACM database, 49 were from the Google Scholar database, 28 were from the IEEE database, and 179 were from the Scopus. After duplicate removal, we were left with 498 studies.

\subsubsection*{Step 3: Select primary studies}
 
In this step, we examined the titles and abstracts of the 498 identified studies and excluded 430 that we deemed irrelevant to our SLR goals. We then applied our five exclusion criteria (i.e.,  EC1-EC5) to the remaining 68 studies to exclude 36. The number of studies filtered based on EC1-EC5 is 8, 6, 3, 12, and 7, respectively.

\vspace{-4pt}
\subsubsection*{Step 4: Forward and backward snowballing}
Backward Snowballing is the process of examining the references of a publication to select additional relevant studies. Using Backward Snowballing, we identified six additional studies. Forward Snowballing is identifying relevant publications that cited our selected preliminary studies. In this step, we found two additional relevant studies. At the end of this step, we had 40 studies, among which eight came from the snowballing steps.

\vspace{-4pt}
\subsubsection*{Step 5: Include additional studies based on expert recommendation}
 In this step, we decided to validate our list of 40 studies based on the opinions of expert researchers. We selected three expert researchers whose names appear in more than three studies. We sent them emails mentioning this SLR's goal and providing the list of 40 papers. We explicitly solicited their suggestions regarding studies missing from our list. Two of the three expert researchers replied to our email, and we included five more studies based on their suggestions.

\subsubsection*{Step 6: Prepare final selection based on inclusion/ exclusion criteria}
In this step, we carefully reviewed the 45 publications and applied EC6,  EC7, and EC8 (i.e., we discarded studies that did not provide recommendations regarding OSS newcomers' task selection or work practices). A total of {16} studies survived this step. While adding five studies in Step 5 based on expert recommendation may question the adequacy of our search strings, we noticed that none of those five studies survived EC6, EC7, and EC8. Our investigation also revealed the lack of `goal' related keywords as the primary reason behind their exclusion by our searches. Our replication package lists the discarded papers with deciding steps~\cite{replication-turzo}. Among the excluded 29 studies, 2 were due to EC7 and  the remaining 27 were due to EC8.

\subsubsection*{Step 7: Data extraction}
In this step, two authors reviewed the 16 papers independently and highlighted sentences suggesting recommendations to newcomers about task selection or work practices. The disagreements between the two authors in identifying the recommendations were resolved in a discussion session. 
We compiled a list of the highlighted sentences in a separate document. We followed a card sorting technique suggested by Zimmermann~\cite{zimmermann2016card} to identify the list of recommendations. We also mapped each of the recommendations to one or more primary studies.

\subsection{RQ1: What are the task-related recommendations for prospective newcomers of 
OSS projects?}
\label{sec:rq1}

\begin{table}
    \centering
      \caption{Research methodology for studies on newcomers' recommendations}
          \vspace{-8pt}
    \resizebox{\linewidth}{!}{
    \begin{tabular}{|p{3cm}|p{0.5cm}|p{4.4cm}|}\hline 
        \textbf{Research method }& \#& \textbf{ Number of studies}  \\ \hline
        Interview / survey &  10 & SLR1, SLR2, SLR4, SLR6, SLR7, SLR8, SLR9, SLR10, SLR11, SLR16 \\ \hline
        Mining software repositories (MSR) & 5 & SLR3, SLR12, SLR13, SLR14, SLR15 \\ \hline        
        Mixed: Survey and MSR &1 &  SLR5\\  \hline       
    \end{tabular}}  
    \label{tab:prior-study}
\end{table}

\begin{table*}[ht!]
	\centering
    \caption{Task-related recommendations for prospective newcomers identified based on our  Systematic Literature Review}
	\label{table:recommendation}
	\resizebox{\linewidth}{!}{
\begin{tabular}{p{0.5cm}|p{4.4cm}|p{7.4cm}|p{4.3cm}}
\hline
\textbf{\#ID} & \textbf{Recommendation}   & \textbf{Rationale}   &      \textbf{Studies} \\ \hline
\multicolumn{4}{l}{ \textbf{Recommendation group: Task selection}}\\ \hline
R1          & Work on smaller tasks first and later choose larger tasks & Working on smaller tasks might increase the likelihood of newcomers' first patch getting accepted  & SLR1\cite{balali2018newcomers}, SLR2\cite{balali2020recommending}, SLR3\cite{huang2021characterizing}, SLR5\cite{rehman2022newcomer}, SLR6\cite{steinmacher2021being}, SLR8\cite{steinmacher2016overcoming}, SLR11\cite{steinmacher2018let}, SLR12\cite{subramanian2020analyzing},
SLR15\cite{subramanian2020empirical} \\ \hline
R2          & Work on Good First Issues (GFIs) (for GitHub projects) &  GFIs are the tasks that are intended to be solved by newcomers. GFIs are less complex, have lower knowledge requirements, are sought by project maintainers, and therefore have higher odds of acceptance & SLR3\cite{huang2021characterizing}, SLR13\cite{tan2020first}, SLR14\cite{xiao2022recommending} \\ \hline
R3          & Add new functionality to the project as your first contribution & If a newcomer adds new functionality, that is currently missing in the project, project maintainers may be more interested in accepting the contribution & SLR2\cite{balali2020recommending},  SLR5\cite{rehman2022newcomer}, SLR6\cite{steinmacher2021being}, SLR12\cite{subramanian2020analyzing}, SLR14\cite{xiao2022recommending} \\ \hline
R4          & Start your contribution by working on minor bug fixes  &  Minor bug fixes have lower knowledge requirements and also help improve the quality of a project  & SLR3\cite{huang2021characterizing}, SLR5\cite{rehman2022newcomer}, SLR12\cite{subramanian2020analyzing}, SLR14\cite{xiao2022recommending},
SLR15\cite{subramanian2020empirical}\\ \hline
R5          & Work on changes where the number of files involved  are few & Task complexity, knowledge requirements, and potential to introduce unintended bugs increase with the number of files involved  & SLR3\cite{huang2021characterizing}, SLR13\cite{tan2020first}  \\ \hline

R6          & Start your contribution by performing less complex tasks & Complex tasks are difficult for newcomers to solve and can possibly demotivate newcomers due to rejections  & SLR1\cite{balali2018newcomers}, SLR3\cite{huang2021characterizing}, SLR4\cite{guizani2021long}, SLR7\cite{santos2022choose}, SLR8\cite{steinmacher2016overcoming}, SLR9\cite{steinmacher2014preliminary}, SLR10\cite{steinmacher2015social}, SLR13\cite{tan2020first}, SLR14\cite{xiao2022recommending}, SLR16\cite{steinmacher2015fostering}  \\ \hline
R7          & Start your contribution by working on documentation changes  &  Documentation changes are less complex, have fewer contribution barriers such as setting up environment or testing, and therefore have a higher odds of getting accepted & SLR3\cite{huang2021characterizing}, SLR5\cite{rehman2022newcomer}, SLR12\cite{subramanian2020analyzing}, SLR14\cite{xiao2022recommending},
SLR15\cite{subramanian2020empirical}\\ \hline
R8          & Work on changes where fewer lines need to be added & Inserting more lines indicate the task might be difficult to solve for newcomers  & SLR3\cite{huang2021characterizing}, SLR13\cite{tan2020first}  \\ \hline
R9          & Work on changes where fewer lines need to be deleted & Before deleting existing lines, a contributor must check for potential usages of those lines and unintended consequences. Therefore, deleting lines is more complex than insertions. & SLR3\cite{huang2021characterizing}, SLR13\cite{tan2020first}  \\ \hline

\multicolumn{4}{l}{ \textbf{Recommendation group: Task preparation}}\\ \hline
R10          & Provide a longer title to help understand the purpose of the change & GFIs with longer titles are positively correlated with acceptance of first changes.  & SLR13\cite{tan2020first}, SLR14\cite{xiao2022recommending}  \\ \hline
R11          & Provide a longer description with detailed information about the task    & GFIs with  longer descriptions are positively correlated with acceptance of first changes & SLR13\cite{tan2020first},  SLR14\cite{xiao2022recommending}  \\ \hline
R12          & Make the title easy to understand  & Well defined titles help reviewers understand the goal of a change  & SLR3\cite{huang2021characterizing}, SLR13\cite{tan2020first},  SLR14\cite{xiao2022recommending}  \\ \hline
R13          & Make the description easy to understand  & If the description text is sufficiently readable, then it would help reviewers understand the submitted changes and might impact acceptance  & SLR3\cite{huang2021characterizing}, SLR13\cite{tan2020first},  SLR14\cite{xiao2022recommending}  \\ \hline

\multicolumn{4}{l}{ \textbf{Recommendation group: Task completion}}\\ \hline
R14          & Be proactive by quickly responding to questions or suggestions  & Proactivity not only shows high interest but also indicates reliability, which is a desired trait of long-term contributors.  & SLR1\cite{balali2018newcomers}, SLR4\cite{guizani2021long}, SLR8\cite{steinmacher2016overcoming}, SLR9\cite{steinmacher2014preliminary}, SLR10\cite{steinmacher2015social}, SLR11\cite{steinmacher2018let}  \\ \hline
R15         & Be patient and focus on addressing review suggestions  
 &  Patience is necessary to go through rigorous quality assurance and iterative improvements to ensure your code meets the project's standards. & SLR1\cite{balali2018newcomers}, SLR8\cite{steinmacher2016overcoming}, SLR9\cite{steinmacher2014preliminary}, SLR10\cite{steinmacher2015social} \\ \hline

\end{tabular}
}%
\end{table*}

Table~\ref{tab:prior-study} shows an overview of the research method from the 16 selected primary studies. Ten out of the 16 studies are qualitative, and most of those recommendations did not include any quantitative evaluation of those recommendations. The five quantitative studies are recent works on characterizing GFIs on GitHub but lack analysis of non-GFIs.
This result further shows the need for a large-scale quantitative analysis.
Table~\ref{table:recommendation} shows our list of 15 recommendations identified based on our SLR. We divided the recommendations into three groups: \textit{Task selection}, \textit{Task preparation}, and \textit{Task completion}. Under the task selection group, we got nine recommendations. The two most popular recommendations within this group for newcomers are to work on less complex tasks (R6), found in 9 primary studies, and to work on smaller tasks (R1), found in 8 studies. Within the \textit{Task preparation} group, we obtained four recommendations, where making the title and description comprehensible ranks first with three studies. We found two \textit{Task completion} recommendations (i.e., what to do after patch submission). While most recommendations are congruent, we found a conflict between R3 and R4, i.e., starting with new features vs. bug fixes.

\vspace{-12pt}
\begin{boxedtext}
\textbf{Key takeaway 1:} \emph{Our SLR identified a total of 15 task-related recommendations for newcomers, where nine are Task selection-related ones. Four recommendations are for \textit{Task preparation}, and the remaining two are for \textit{Task completion}. }
\end{boxedtext}
\section{Stage 2: Empirical evaluation of newcomers' onboarding recommendations}
\label{sec:stage2}
The following subsections detail our research methodology to empirically evaluate the task-related recommendations identified based on our SLR, as shown in Table~\ref{table:recommendation},  and investigation results.

\subsection{Research Method}
We followed a six-step method to answer RQ2 and RQ3.  In the first step, we defined one quantitative metric to measure a code change's compliance against each recommendation listed in Table~\ref{table:recommendation}. Second, we identified a set of projects to test each recommendation-based hypothesis (i.e., whether a recommendation has a positive or negative association with code acceptance). Third, we mine code review /pull request datasets for the selected projects. Fourth, we computed the attributes identified in the first step using our dataset. Fifth, we trained and evaluated logistic regression models for each project /project group following Harrell Jr.'s guidelines~\cite{harrell2015regression}. 
We trained two models for each dataset, one for newcomers (RQ2) and the other for non-newcomers (RQ3).
 We test each attribute's significance using the Wald Chi-Square statistics (Wald $\chi^2$) test and direction of influence using the Odds Ratio (OR).  Finally, we map the results of our regression models back to our recommendations to provide verdicts on whether our analyses support, contradict, or are inconclusive in each recommendation. The following subsections detail these six steps.

\begin{table*}
	\centering
    \caption{Quantitative metrics selected for recommendations obtained based on SLR}  \label{table:patch_character}
	
\resizebox{\linewidth}{!}{
\begin{tabular}{p{2.5cm} p{6cm} p{6cm} p{1.4cm}}
\hline
\multicolumn{1}{c}{\textbf{Metric}} & \multicolumn{1}{c}{\textbf{Description}}                                                                   & \multicolumn{1}{c}{\textbf{Rationale}} & \multicolumn{1}{c}{\textbf{Reco(s).}} \\ \hline
\rowcolor[HTML]{EFEFEF} 
Total churn (churn)                            & Total number of lines that have been changed (total added and deleted lines) & Complexity of changes increase with total churn required for a task  & R1 \\

Is Good first issue (GFI)*?                               & Whether the associated issue had one of the GFI tags on GitHub.    &  Many projects use GFI tags to encourage prospective newcomers' onboarding.     &  R2 \\

     \rowcolor[HTML]{EFEFEF} 
New feature                         & Does the change add a new feature or enhancement?  & New feature-related tags or a higher number of new files indicates the inclusion of new features that may be currently missing. &  R3  \\

Is bug fix? (bug fix)                             & Whether the change under review aims to fix an existing bug   &  Prior studies recommended newcomers to work on smaller bug fixing and `bug fix' indicates whether the submitted change is for bug fixing or not. &  R4 \\

\rowcolor[HTML]{EFEFEF} 
The number of files (file)                            & Total number of files associated with the change under review  &   Changes requiring multiple file modifications may be difficult to comprehend for newcomers. & R5  \\

Change entropy (entropy)                        & CE measure the dispersion of a change among the associated files. If $n$ is the number of files in a change, $r_k$ is the ratio of change for the file $k$, then $entropy = -\frac{\sum_{k=1}^{n} (r_k \times log_2 r_k )}{log_{2}(n+1)}$   &    
Following the recommendation of prior studies~\cite{hassan2009predicting, thongtanunam2017review}, we use the change entropy measure to compute the complexity of a task. 
      &    R6 \\

\rowcolor[HTML]{EFEFEF}
Doc file ratio (doc file)                   & Documentation file ratio is the ratio between the number of documentation files and the total number of files involved. & Higher `doc file' indicates the change under review includes mostly documentation changes & R7 \\

Insertions (insert)                             & Total number of lines inserted into the files under review.   &  Number of lines inserted is correlated with the complexity of a task &  R8   \\

\rowcolor[HTML]{EFEFEF} 
Deletions (delete)              & Total number of lines deleted from the files under review                & Number of lines deleted  are also correlated with task complexity &  R9  \\ 
 
Length of title (title length)   & Number of words in the title    &  A meaningful and detailed title helps to draw the attention of the project maintainers to look into change  & R10  \\

\rowcolor[HTML]{EFEFEF} 
Length  of description (desc. length)   & Number of words in the description text    &   An elaborated description helps the project maintainers to understand the change's context, rationale, and solution strategy & R11  \\ 

Title readability (title read.)    & The Flesch reading ease\cite{kincaid1975derivation} index of the title  &  An easy to understand title helps project maintainers' comprehension & R12  \\ 

\rowcolor[HTML]{EFEFEF} 
Description readability (desc read.)    & The Flesch reading ease of the description text  &  An easy to understand description helps project maintainers' comprehension & R13  \\

Promptness  & We define the time interval between the review of a patchset and the next patchset submission as author delay. We calculate `promptness' as the inverse of author delays.  &  A lower author delay indicates higher proactiveness of the author to address suggested changes. & R14  \\

\rowcolor[HTML]{EFEFEF} 
Review iterations (iteration)                         & The total number of patchsets submitted for review.   &  Authors submit new patches to add changes based on reviewers' suggestions. A higher number of iterations indicates higher efforts during a review. & R15   \\
\hline

\multicolumn{4}{l}{ \small{*GitHub dataset only}}

\end{tabular}
}%
\end{table*}

\subsubsection{Recommendation to quantitative metrics mapping}
We select a quantitative metric for each recommendation based on prior works on code reviews~\cite{thongtanunam2017review,mcintosh2014impact,mcintosh2016empirical,sultana2022code,paul2021security}, and the 16 newcomers' onboarding-related studies found in our SLR.
Identifying metrics for R2,   R4, R5, and R8-R11 was straightforward, as those directly allude to measures. 
While \textit{ task size (R1)} can be measured using various metrics, we use `total churn', the total number of lines added, modified, or deleted in a change, since this metric has been used by prior code review studies to estimate change size~\cite{thongtanunam2017review,mcintosh2016empirical}. 
In addition to new feature-related labels, we map \textit{adding new functionality (R3)} to whether a new file was added since a significant share of pull requests do not have any label.  
We compute the `Change entropy' metrics introduced by Hassan~\cite{hassan2009predicting} to measure \textit{task complexity (R6)}, as this metric has also been used in several code review-related studies to compute change complexity~\cite{thongtanunam2017review, mcintosh2016empirical}.
To map \textit{R7: working on documentation}, we compute the ratio of files in the change with documentation-related extensions. 
To map \textit{R12: readable titles} and \textit{R13: descriptions}, we select Flesch Reading ease~\cite{kincaid1975derivation}, since it is one of the most widely used readability scores and prior research has found this metric suitable to assess readabilities of technical manuals~\cite{kincaid1988electronic}.
We map \textit{R14: authors'  proactivity} with author promptness (i.e., how quickly the author resubmitted after a review) and \textit{R15: being patient} with the number of iterations (i.e., how many times the author resubmitted the change to address suggested changes).
Table~\ref{table:patch_character} shows the mapping among the 15 recommendations and the selected metrics.

\subsubsection{Sample Selection}
\label{sec:dataset}
For this study, we selected projects from two sources. First, we selected five projects using Gerrit-based code reviews to accept newcomers' code changes. Gerrit is among the most popular code review tools many non-GitHub-based OSS projects use. Our selection criteria for Gerrit projects include i) Has been using Gerrit for at least five years, ii) Has at least 50,000 CRs, and iii) Has at least 500 newcomers.
We selected these five criteria to ensure adequate newcomers' CRs, which is a requisite to train reliable regression models. We identified a total of 21 publicly available Gerrit-managed CR repositories~\cite{bosu2013modeling}, where six satisfied our selection criteria. Between the two Google-managed projects, we selected Android instead of Chromium OS since it had a higher number of newcomers during our study period.  We selected all four non-Google projects satisfying our criteria. Therefore, our selection of Gerrit-based OSS projects are i) Android: a mobile operating system sponsored by Google; ii) LibreOffice: an office development platform sponsored by a nonprofit foundation; iii) Typo3: a web content management system run by a nonprofit community; iv) oVirt: a platform for managing virtual machines, managed by a for-profit company; and v) Wikimedia:  wiki management software, community-based. These five projects represent diverse domains, sponsorship, and programming languages. These projects have also been subject to prior code review studies \cite{bosu2019diversity,sultana2022code}.
Next, we selected OSS projects using the PR-based development model~\cite{gousios2014exploratory} to accept newcomers' code changes.  We use a GitHub project search tool developed by Dabic \emph{et} al.~\cite{Dabic:msr2021data} for this goal.
 Following the recommendations of Kalliamvakou \textit{et} al.~\cite{kalliamvakou2016depth}, we excluded projects that do not meet the following four criteria. 

\begin{enumerate}[leftmargin=*]
  \item Uses one of the top ten programming languages on GitHub:  Java, C, C++, Python, JavaScript, C{\#}, Go, PHP, Typescript, and Ruby;
  \item The source code is publicly available and has at least 20 contributors.
  \item Has at least 20 PRs during the last three months, according to our dataset.
  \item Has at least five stars.
\end{enumerate}

The first three criteria ensure the selection of OSS projects with adequate contributors and development activities. 
The remaining criterion reduces the search space; without this filter, the number of projects grows exponentially but adds only trivial OSS projects.

Our search found 61,123 projects. We divided the surviving projects into three groups based on development activity:  i) GitHub (L): Low development activity with $<$ 2 per week, i.e., eight pull requests per month; ii) GitHub (M): Medium development activity with between 8 to 32 pull requests per month, and iii) GitHub (H): High development activity with $>$ 32 pull requests per month. We computed these thresholds based on distributions of PR/month. Approximately $>$32 PR/month represents the top 10\%, and  $>$8 PR/month represents the top 75\%. We took PR/month as our selection criteria since we noticed that many projects on GitHub have a very low number of PRs despite having many commits and unique contributors, as these projects use GitHub as a mirror and conduct most development activities on another platform. We randomly selected 1,155 projects, where exactly 385 were selected from each group to achieve 95\%  confidence interval and 5\%  margin of error~\cite{cochran1977sampling} considering the population is all the available projects in GitHub. Our sample includes popular OSS projects such as Kubernetes, rust, OpenCart, rails, pandas, and Odoo.

\subsubsection{Dataset Preparation}
Table~\ref{tab:project-overview} shows an overview of our dataset.
\textit{Gerrit data:} We have used a Gerrit Miner tool \cite{bosu2013impact} to collect
all the submitted review requests till 30th April, 2023.
Our Gerrit dataset includes 1,438,042 code review (CR) requests from the five selected projects.
We adopted a keyword-based filtering approach proposed by Bosu \textit{et al.}~\cite{bosu2019diversity} to identify and exclude bot-generated interactions. With this approach, accounts with the following set of keywords were marked:   `bot’, `CI’, `Jenkins’, `build’, `auto’, `devop', `hook', `workflow', and ‘travis’. Next, we manually inspected each marked account to make a final determination.
We leverage the Active Learning Fingerprint Based Anti-Aliasing (ALFAA) approach proposed by Amreen et al.~\cite{amreen2020alfaa} to identify multiple Gerrit accounts belonging to the same person into one account.

\vspace{4pt}
\noindent\textit{GitHub:}
We wrote Python scripts using the PyGitHub library~\cite{pygithub} to mine all the pull requests for the selected projects directly from GitHub. Our GitHub dataset includes a total of 5,996,077 pull requests (PR).

\begin{table}[]
    \caption{Overview of the selected dataset}
    \centering
    \begin{tabular}{|l|r|r|} \hline
    \textbf{Dataset} & \textbf{\# CR /PR} & \textbf{\# of newcomer} \\ \hline
    Android     &   637,064 & 7,683 \\ \hline
    LibreOffice   &127,812  & 1,223 \\ \hline
    oVirt & 113,465 & 517\\ \hline
    Typo3 &  65,399 & 978\\ \hline
    Wikimedia & 494,302  &1,854\\ \hline \hline
    GitHub (L)   &  140,376 & 23,259 \\ \hline
    GitHub (M)  & 695,812 &  78,894 \\ \hline
    GitHub (H)  & 5,159,889 & 262,013\\ \hline
    \end{tabular}
    \label{tab:project-overview}
    \vspace{-12pt}
\end{table}

\subsubsection{Attribute Calculation}
We wrote Python scripts and MySQL queries to compute the metrics listed in Table~\ref{table:patch_character}. However, we had to define additional heuristics for the following set of metrics.
\begin{itemize}[leftmargin=*]
    \item \textit{Is newcomer:} Following the recommendation of Steinmacher \textit{et al.}, we consider a contributor as a newcomer to a project till he/ she/ they have got their first CR /PR accepted~\cite{steinmacher2015social,subramanian2020analyzing}. 
     We consider all other contributors as non-newcomers and use them as the `control group'. However, we acknowledge that non-newcomers are diverse and can be further divided into several sub-categories~\cite{lee2017understanding,pinto2016more,pham2013building,barcomb2020managing}.  
    \item \textit{GFI:} To identify GFIs, we use issue labels listed by Tan \textit{et al.}~\cite{tan2020first} and in a newcomer contribution guideline~\cite{awesome-newcomer}. 
    \item We extracted extensions for each file and marked a file as doc if its extension belonged to one of the documentation-indicative types (e.g., `log', `md', `rst', `txt',  `install',`release', `wiki' and `LICENCE')
    \item We use the \code{textstat} library to compute readability scores.
    \item Similar to prior studies~\cite{mcintosh2014impact,thongtanunam2017review,sultana2022code}, we mark a CR/PR as a bug fix if it includes one of the bug-related labels (e.g., `bug',  and `defect') or the description includes bug associated words such as `bug', `fix, `defect', and `Bug Id:'.
\end{itemize}

\subsubsection{Model Training and Evaluation}
\label{model-training}
We train Multivariate Logistic Regression (MLR) models to identify associations between attributes listed in Table~\ref{table:patch_character} and CR/PR acceptance. 
Although MLR has been recently used in Machine learning, it has been used for decades by statisticians since its invention by David Cox in 1958~\cite{cox1958regression}. MLR effectively investigates the relationships between a binary dependent variable and one or multiple independents \cite{bewick2005statistics}. There are two categories of MLR models: i) \textit{predictive} -- where the goal is to predict future outcomes based on past data, and ii) \textit{inferential} -- where the goal is to identify associations between the dependent and multiple independents. Predictive models are evaluated using precision, recall, F-score, and accuracy. On the other hand, inferential models are evaluated using their ability to explain observed variances, usually measured using Pseudo-$R^2$ ~\cite{harrell2015regression}.  
Based on our study's goal, we train inferential MLR models.

Multicollinearity arises when two or more independent variables are highly correlated and is a threat to inferential MLR models~\cite{mansfield1982detecting}. 
We followed the approach of Harell Jr. \cite{harrell2015regression}, which has been used in many prior SE studies~\cite{thongtanunam2017review,mcintosh2014impact,paul2021security}. Our models use the 15 factors listed in Table \ref{table:patch_character} as the independent variables and acceptance of PR/CR as the dependent variable. 
We used 14 independents for Gerrit projects, as GFI tags are not used among Gerrit-based projects.  We named our dependent variable as \emph{is\_accepted}, set to `1' if the patch is accepted and `0' otherwise. While computing \emph{is\_accepted} was straightforward for Gerrit projects, this variable may be incorrect on Github~\cite{gousios2014exploratory,zhou2019fork}. 
We followed the heuristics suggested by Gousios \textit{et al.} \cite{gousios2014exploratory} to resolve potentially incorrect acceptance status on GitHub.
For each of the eight datasets (i.e., 5 Gerrit and 3 GitHub), we trained two MLR models. The first model is trained using the CRs / PRs submitted by newcomers to assess RQ2. The second model is trained using non-newcomers'  PRs /CRs to answer RQ3. We log-transformed four variables (i.e., insert, delete, delay, and churn) as those had highly skewed distributions.
Our model training and evaluation steps are as follows.

\vspace{4pt}
\noindent\textit{Step 1: Correlation and Redundancy Analysis:} 
To identify highly correlated explanatory variables (i.e., Multicollinearity), we applied Sarle's Variable Clustering (VURCLUS) approach \cite{sarle1990sas}. We used Spearman's rank-order correlation test \cite{statistics2013spearman} to build a hierarchical cluster representation of explanatory variables. Following the suggestion of Hinkle \textit{et al.} \cite{hinkle1998applied}, we used the correlation coefficient $|\rho| = 0.7$ as the cutoff threshold. If multiple explanatory variables belong to the same cluster, we picked only the one with the highest correlation with the dependent. For example, in most models, we found that log(code churn) and log (insert) are highly correlated (i.e., $|\rho| \geq 0.7$). Since log(insert) has a higher correlation with the dependent,  we dropped log(code churn). Even if not highly correlated, explanatory variables can still be redundant, eventually affecting the model's performance. We use the \code{redun} function of the \code{rms} R package with the threshold $R^2 \geq$  0.9  to identify potential redundant factors among the ones surviving the multicollinearity threshold.  


\vspace{4pt}
\noindent \textit{Step 2: Model evaluation: }
In this study, we use Veall and Zimmerman Pseudo  $R^2$~\cite{veall1994evaluating} to assess the goodness of fit of the MLR models since prior research~\cite{smith2013comparison} found this measure having closer correspondence to ordinary least square $R^2$. The Pseudo ${R^2}$ rows in Tables~\ref{table:gerrit_result_newcomer}, ~\ref{table:github_result-nc}, ~\ref{table:gerrit_result_nonnewcomer}, and ~\ref{table:github_result-non-nc} represent goodness of fit for the models. We also conducted Log-likelihood tests to assess each model against a null model. Our models significantly differ from a null model and can be reliably used for inferential statistics.

\vspace{4pt}
\noindent \textit{Step3: Interpretation of results:} 
To determine the associations of these factors with CR /PR acceptance, we compute odds ratios (OR). OR shows how the dependent variable changes if a factor changes by one unit while all other factors remain constant.  OR $>$ 1 implies that the dependent variable has greater odds of occurring as that particular factor increases, whereas OR $<$ 1 suggests the opposite.
If a factor is dichotomous (e.g., bug fix), OR indicates how the odds of acceptance change if the factor is present in contrast to its absence (e.g., not a bug fix), given the remaining factors are identical.

\vspace{4pt}
\noindent \textit{Power of each variable:} 
 We conducted an ablation study to understand how the inclusion/exclusion of each independent variable influences a model's association with the dependent. In this design, we start with a null model and compute its unexplained variance (i.e., \textit{Null deviance ($DIV_{null}$)})\footnote{Deviance is a measure to assess goodness of fit for regression models. It is computed based on the difference between trained and perfect models. The lower the deviance, the better the model.}. Then, we measure the change in residual deviance ($\Delta DIV_{residual}$) with the addition of each independent. Therefore, the explanatory power of an independent is computed according to the following equation: $EP$  = $\frac{\Delta DIV_{residual}}{DIV_{null}} * 100$. The higher the value of \textit{EP(\%)}, the better that attribute explains the dependent variable.

\begin{table*}
	\centering
    \caption{Results of logistic regression models for the five Gerrit-based projects for newcomers. $OR^p$ indicates the Odds ratio for the attribute with a marker indicating statistical significance (i.e., $p$). $EP(\%)$$\rightarrow$percentage of deviance reduced by that attribute. Cells in \positive{ light blue} backgrounds represent positive associations with the attribute, and cells in \negative{ light pink} backgrounds represent negatives. The same interpretation rules also apply to Tables   ~\ref{table:github_result-nc}, ~\ref{table:gerrit_result_nonnewcomer}, and ~\ref{table:github_result-non-nc}.}
    \vspace{-6pt}
	\label{table:gerrit_result_newcomer}
	 \resizebox{\linewidth}{!}{\begin{tabular}{|l|l|r|r|r|r|r|r|r|r|r|r|}
\hline
\multicolumn{2}{|c|}{\textbf{Project} $\rightarrow$}  & \multicolumn{2}{c|}{\textbf{Android}}  & \multicolumn{2}{c|}{\textbf{LibreOffice}}  & \multicolumn{2}{c|}{\textbf{oVirt}} & \multicolumn{2}{c|}{\textbf{Typo3}} & \multicolumn{2}{c|}{\textbf{Wikimedia}}  \\ \hhline{------------}
\multicolumn{2}{|c|}{\textbf{Pseudo $R^{2} \rightarrow$}} &  \multicolumn{2}{c|}{0.177$^{***}$} & \multicolumn{2}{c|}{0.273$^{***}$} & \multicolumn{2}{c|}{0.388$^{***}$} & \multicolumn{2}{c|}{0.262$^{***}$} & \multicolumn{2}{c|}{0.054$^{***}$} \\ \hline  \hline

 \textbf{Attribute} & \textbf{Reco(s)} & \textbf{OR$^{p}$}  & \multicolumn{1}{c|}{\textbf{\textit{EP(\%)}}} & 
 \textbf{OR$^{p}$} & \multicolumn{1}{c|}{\textbf{\textit{EP(\%)}}} & 
 \textbf{OR$^{p}$} & \multicolumn{1}{c|}{\textbf{\textit{EP(\%)}}} & 
 \textbf{OR$^{p}$} & \multicolumn{1}{c|}{\textbf{\textit{EP(\%)}}} & \textbf{OR$^{p}$} & \multicolumn{1}{c|}{\textbf{\textit{EP(\%)}}} \\ \hline

  log(code churn) & R1 & -- & -- & -- & -- & -- & -- & -- & -- & -- & --      \\ \hline                                                         
 new feature & R3 &  \positive{  1.001$^{*}$ } & 0.06& 1.00 & 0.00& 1.08 & 0.96& 1.01 & 0.32& 1.00 & 0.04   \\ \hline                                   
  bugfix & R4 & \positive{  2.41$^{***}$ } & 4.47& 0.90 & 0.25& \positive{  2.64$^{***}$ } & 4.67& \positive{  1.59$^{**}$ } & 0.50& \positive{  1.44$^{***}$ } & 0.76           \\ \hline      
   file count & R5 & -- & -- & -- & -- & -- & -- & -- & -- & -- & --         \\ \hline                                                          
  entropy & R6 &0.89 & 0.01& 0.98 & 0.00& \negative{  0.51$^{*}$ } & 0.47& 1.37 & 0.09& 0.93 & 0.00         \\ \hline                            
  doc file & R7 & \positive{  2.27$^{***}$ } & 0.27& \positive{  4.56$^{*}$ } & 0.28& 0.82 & 0.00& 0.73 & 0.11& 1.20 & 0.04       \\ \hline                      
 log(insert) & R8  & \negative{  0.89$^{***}$ } & 0.10& \negative{  0.90$^{**}$ } & 0.03& \negative{  0.66$^{***}$ } & 0.49& \negative{  0.86$^{**}$ } & 0.29& 0.96 & 0.00             \\ \hline
  log(delete) & R9 &  \positive{  1.05$^{***}$ } & 0.02& 1.00 & 0.00& \positive{  1.16$^{*}$ } & 0.03& 1.09 & 0.17& 0.99 & 0.00        \\ \hline                      
  title length & R10 & \negative{  0.999$^{***}$ } & 0.13& 1.00 & 0.09& 1.00 & 0.06& 1.01 & 0.06& 1.00 & 0.00        \\ \hline                             
 desc. length & R11 & \negative{  0.999$^{*}$ } & 0.03& 1.00 & 0.08& 1.00 & 0.04& \negative{  0.999$^{**}$ } & 0.54& 1.00 & 0.06    \\ \hline                      
 title read. & R12 & 1.00 & 0.01& 1.00 & 0.00 & -- & --& 1.00 & 0.00 & -- & --             \\ \hline                                   
 desc read. & R13 & 1.00 & 0.00 & -- & --& 1.00 & 0.06& 1.00 & 0.11& 1.00 & 0.01                                     \\ \hline
promptness & R14 & \negative{  0.94$^{***}$ } & 0.08& \positive{  1.76$^{***}$ } & 2.29& 1.10 & 0.12& \positive{  1.59$^{***}$ } & 2.59& \positive{  1.25$^{***}$ } & 0.83               \\ \hline
 iteration & R15 &  \positive{  1.17$^{***}$ } & 3.09& \positive{  1.80$^{***}$ } & 11.19& \positive{  1.51$^{***}$ } & 14.19& \positive{  1.41$^{***}$ } & 8.83& \positive{  1.05$^{***}$ } & 0.61 \\ \hline

\multicolumn{12}{p{14cm}}{ *** , **, and *  represent statistical significance at $p <$ 0.001, $p <$ 0.01, and $p <$ 0.05 respectively. No markers for $p$  indicate statistical insignificance. $-$ indicates attribute dropped from the model due to multicollinearity.}
\end{tabular}%
}
\end{table*}

\begin{table*}
	\centering
    \caption{Results of logistic regression models for Github newcomers }
	\label{table:github_result-nc}
\begin{tabular}{|l|l|r|r|r|r|r|r|}
\hline
\multicolumn{2}{|c|}{\textbf{Dataset} $\rightarrow$}  & \multicolumn{2}{c|}{\textbf{GitHub(L)}}  & \multicolumn{2}{c|}{\textbf{GitHub(M)}}  & \multicolumn{2}{c|}{\textbf{GitHub(H)}} \\ \hline
 \multicolumn{2}{|c|}{\textbf{Pseudo $R^{2} \rightarrow$}} & \multicolumn{2}{c|}{0.094$^{***}$} &  \multicolumn{2}{c|}{0.092$^{***}$} &  \multicolumn{2}{c|}{0.127$^{***}$} \\ \hline

 \textbf{Attribute} & \textbf{Reco(s)} & \textbf{OR$^{p}$} & \textbf{\textit{EP(\%)}}  &  \textbf{OR$^{p}$} & \textbf{\textit{EP(\%)}}  & \textbf{OR$^{p}$} & \textbf{\textit{EP(\%)}} \\ \hline

  log (code churn) & R1 & -- & -- & -- & -- & -- & --    \\ \hline                                
  GFI & R2 &1.42 & 0.00& \positive{  2.02$^{**}$ } & 0.01& \positive{  5.57$^{***}$ } & 0.02\\ \hline                  
  new feature & R3 &1.00 & 0.01& \negative{  $0.999^{***}$ } & 0.06& \negative{  $0.999^{***}$ } & 0.08      \\ \hline                                
  bug fix & R4 & \positive{  1.27$^{***}$ } & 0.41& \positive{  1.34$^{***}$ } & 0.40& \positive{  1.18$^{***}$ } & 0.08     \\ \hline  
  file count &R5 & -- & -- & -- & -- & -- & --    \\ \hline         
  entropy & R6 & \negative{  0.82$^{***}$ } & 0.05& \negative{  0.80$^{***}$ } & 0.07& \negative{  0.85$^{***}$ } & 0.04 \\ \hline 
  doc file & R7 & \positive{  5.45$^{***}$ } & 1.22& \positive{  4.08$^{***}$ } & 0.98& \positive{  5.39$^{***}$ } & 1.82 \\ \hline
  log(insert) & R8 & \negative{  0.87$^{***}$ } & 2.76& \negative{  0.91$^{***}$ } & 2.56& \negative{  0.91$^{***}$ } & 2.92 \\ \hline
  log(delete) & R9 & \negative{  0.92$^{***}$ } & 0.33& \negative{  0.88$^{***}$ } & 0.65& \negative{  0.85$^{***}$ } & 1.09  \\ \hline
  title length & R10 & \negative{  $0.999^{***}$ } & 0.06& \negative{  0.999$^{**}$ } & 0.01& \negative{  $0.999^{***}$ } & 0.01   \\ \hline
  desc. length & R11 & \negative{  $0.999^{***}$ } & 0.28& \negative{  $0.999^{***}$ } & 0.01& \positive{  1.001$^{***}$ } & 0.22 \\ \hline                                 
  title read. & R12 & 1.00 & 0.00& 1.00 & 0.00& \negative{  $0.999^{***}$ } & 0.03 \\ \hline                                  
 desc read. & R13 & \negative{  $0.999^{***}$ } & 0.05& \negative{  $0.999^{***}$ } & 0.05& \negative{  $0.999^{***}$ } & 0.02 \\ \hline                            
  promptness & R14 & 1.00 & 0.00& \positive{  1.01$^{***}$ } & 0.05& \negative{  0.99$^{***}$ } & 0.01          \\ \hline  
  iteration & R15 & 1.01 & 0.00& \positive{  1.03$^{***}$ } & 0.04& \negative{  0.99$^{***}$ } & 0.01        \\ \hline

\multicolumn{8}{p{11cm}}{ *** , **, and *  represent statistical significance at $p <$ 0.001, $p <$ 0.01, and $p <$ 0.05 respectively. No markers for $p$  indicate statistical insignificance.}
\end{tabular}%
\end{table*}

\subsection{RQ2: How do compliance with OSS newcomers' onboarding recommendations affect the odds of acceptance of their first patches?}
\label{sec:rq2}
Tables \ref{table:gerrit_result_newcomer} and \ref{table:github_result-nc} show the results of our regression models for newcomers in Gerrit and GitHub projects. Two of the 15 variables are missing from these tables, as \textit{log(code churn)} is redundant with  \textit{log(insert)} and \textit{num file} is redundant with \textit{entropy} for all eight models. 

For a project, the  $OR^p$ columns in Tables   \ref{table:gerrit_result_newcomer} and \ref{table:github_result-nc} list effect sizes for the attributes, estimated using \textit{Odds ratio}.  
 For example,  in the LibreOffice project, the $OR^p$ value for the newcomer group for the \textit{doc file} factor is 4.56 (Table \ref{table:gerrit_result_newcomer}). This result indicates that if a newcomer's change comprises 100\% doc files in LibreOffice, the odds of acceptance are 4.56 times higher than that of newcomers' changes including 0\% doc files, while holding the other variables constant. 
The \textit{EP(\%)} columns in Tables \ref{table:gerrit_result_newcomer} and \ref{table:github_result-nc} indicate the explanatory power of each attribute for the model, with higher \textit{EP(\%)}s indicating better predictors.  For example, in Table \ref{table:gerrit_result_newcomer}, the value of \textit{EP(\%)} for \emph{bugfix} is 4.47\% in Android indicating percentage contribution of that variable for Android's model. 
We consider a recommendation supported only if the mapped factor has significant positive associations for a project. In case of negative associations, the recommendation is contradicted. The recommendation is insignificant if no significant association ($ p<0.05$) is seen. 

None of the recommendations showed significant support across all eight datasets. 
Two recommendations (\textit{R4: bugfix} and \textit{R8: fewer line insertions)} are supported across seven of eight datasets for newcomers.  
We noticed the most contradiction for \textit{R10: title length} and \textit{R11: description length}, with four and five datasets suggesting opposite associations, respectively. However, the explanatory powers of  \textit{title length} and \textit{description length} attributes are low ($<0.2\%$) across all models, indicating these variables have low contributions to overall model performance.
Similarly, R12 and R13 are insignificant across all projects or negatively associated with low explanatory powers. 

\textit{R2: GFI}, which is only applicable for GitHub datasets, has significant positive associations for GitHub (M) and GitHub (H) but is insignificant for GitHub (L). 
\textit{R6: Starting with low complexity changes}, it is supported across four datasets and insignificant for four Gerrit projects except for oVirt. 
\textit{R7: Working on documentation} is supported across five of the eight datasets and is one of the predictors with higher explanatory powers in GitHub. 
We noticed context-dependencies for R3, R9,  R14, and R15. 
While \textit{R3: adding new features} significantly lowers acceptance odds on GitHub (M) and GitHub (H), an opposite association is seen in Android. Similarly,  contradicting \textit{R9: fewer lines deletions}\footnote{Interpretations for R6, R8, and R9 are opposite of the associations with mapped variables since those suggest changes with lower complexities or fewer line insertions /deletions, but the mapped variables measure complexity and the number of lines inserted/deleted\revision{.}}, the odds of acceptance increase with the number of lines deleted in Android and oVirt. However, we notice significant support for R9 across all three GitHub datasets. \textit{R14: being practive} is supported for four datasets (i.e., LibreOffice, Typo3, Wikimedia, and Github (M)), contradicts for two (Android and Github (H)), and insignificant for the remaining two. \textit{R15: being patient} is supported across all Gerrit projects and GitHub (M) but contradicted in GitHub (H). Task completion-related recommendations (i.e., R14 and R15) have high explanatory powers across all datasets and are important predictors.

\vspace{-12pt}
\begin{boxedtext}
\textbf{Key takeaway 2:} \emph{Working on bug fixes, good first issues,  changes requiring fewer added lines, and low complexity have significantly higher odds of acceptance for most Gerrit and GitHub projects. }
\end{boxedtext}

\vspace{-8pt}
\begin{boxedtext}
\textbf{Key takeaway 3:} \emph{Four recommendations (i.e., R3, R9,  R14, and R15) are context-dependent for newcomers, with positive associations in one dataset but the opposite association in another. Therefore, newcomers should also consider project norms and expectations before following them. }
\end{boxedtext}

\begin{table*}
	\centering
    \caption{Results of logistic regression models for the five Gerrit-based projects for non-newcomers. }
	\label{table:gerrit_result_nonnewcomer}
	 \resizebox{\linewidth}{!}{\begin{tabular}{|l|l|r|r|r|r|r|r|r|r|r|r|}
\hline
\multicolumn{2}{|c|}{\textbf{Project} $\rightarrow$}  & \multicolumn{2}{c|}{\textbf{Android}}  & \multicolumn{2}{c|}{\textbf{LibreOffice}}  & \multicolumn{2}{c|}{\textbf{oVirt}} & \multicolumn{2}{c|}{\textbf{Typo3}} & \multicolumn{2}{c|}{\textbf{Wikimedia}}  \\ \hhline{------------}
\multicolumn{2}{|c|}{\textbf{Pseudo $R^{2} \rightarrow$}} &  \multicolumn{2}{c|}{0.134$^{***}$} & \multicolumn{2}{c|}{0.061$^{***}$} & \multicolumn{2}{c|}{0.086$^{***}$} & \multicolumn{2}{c|}{0.099$^{***}$} & \multicolumn{2}{c|}{0.136$^{***}$} \\ \hline  \hline

 \textbf{Attribute} & \textbf{Reco(s)} & \textbf{OR$^{p}$}  & \multicolumn{1}{c|}{\textbf{\textit{EP(\%)}}} & \textbf{OR$^{p}$} & \multicolumn{1}{c|}{\textbf{\textit{EP(\%)}}} & \textbf{OR$^{p}$} & \multicolumn{1}{c|}{\textbf{\textit{EP(\%)}}} & \textbf{OR$^{p}$} & \multicolumn{1}{c|}{\textbf{\textit{EP(\%)}}} & \textbf{OR$^{p}$} & \multicolumn{1}{c|}{\textbf{\textit{EP(\%)}}} \\ \hline

  log (code churn) & R1 & -- & -- & -- & -- & -- & -- & -- & -- & -- & --      \\ \hline                                                       
  new feature & R3 & \positive{  1.001$^{*}$ } & 0.01& 1.00 & 0.00& 1.00 & 0.00& 1.00 & 0.00& 1.00 & 0.01                \\ \hline                     
  bug fix & R4 & \positive{  1.81$^{***}$ } & 0.02& 0.92 & 0.01& \positive{  2.24$^{***}$ } & 1.27& \positive{  1.31$^{**}$ } & 0.03& \positive{  1.09$^{***}$ } & 0.56          \\ \hline     
  file count & R5 & -- & -- & -- & -- & -- & -- & -- & -- & -- & --      \\ \hline                                                           
  entropy & R6 & 1.00 & 0.00& 1.22 & 0.05& \positive{  1.33$^{*}$ } & 0.15& 1.37 & 0.09& 1.24 & 0.05     \\ \hline                              
  doc file & R7 & \positive{  1.27$^{***}$ } & 0.03& \negative{  0.90$^{*}$ } & 0.01& 1.41 & 0.01& 1.35 & 0.07& 0.88 & 0.03      \\ \hline                    
 log(insert) & R8 & \negative{  0.91$^{***}$ } & 1.54& \negative{  0.91$^{**}$ } & 0.24& \negative{  0.84$^{***}$ } & 0.49& \negative{  0.83$^{**}$ } & 0.45& 0.92 & 0.63        \\ \hline   
  log(delete) & R9 & \positive{  1.02$^{***}$ } & 0.01& 0.93 & 0.04& \positive{  1.06$^{*}$ } & 0.12& 1.09 & 0.36& 1.02 & 0.01                    \\ \hline   
  title length & R10 & \negative{  0.999$^{***}$ } & 0.15& 1.00 & 0.07& 1.00 & 0.01& 1.00 & 0.00& 1.00 & 0.04                            \\ \hline       
  desc. length & R11 & \negative{  0.999$^{*}$ } & 0.01& 1.00 & 0.00& 1.00 & 0.00& \negative{  0.999$^{**}$ } & 0.01& 1.00 & 0.03      \\ \hline              
  title read. & R12 & 1.00 & 0.00& 1.00 & 0.13& 1.00 & 0.11& 1.00 & 0.00& 0.99 & 0.37               \\ \hline                 
  desc read. & R13 & 1.00 & 0.00 & -- & --& 1.00 & 0.02& 1.00 & 0.00& 1.00 & 0.76   \\ \hline                                 
  promptness & R14 & \positive{  1.33$^{***}$ } & 6.10& \positive{  1.78$^{***}$ } & 1.64& 1.74 & 2.40& \positive{  1.62$^{***}$ } & 4.61& \positive{  1.97$^{***}$ } & 7.08            \\ \hline 
 iteration & R15 & \positive{  1.07$^{***}$ } & 0.14& \positive{  1.45$^{***}$ } & 1.86& \positive{  1.06$^{***}$ } & 0.37& \positive{  1.16$^{***}$ } & 0.83& \positive{  1.01$^{***}$ } & 0.06 \\ \hline

\multicolumn{12}{p{14cm}}{ *** , **, and *  represent statistical significance at $p <$ 0.001, $p <$ 0.01, and $p <$ 0.05 respectively. No markers for $p$  indicate statistical insignificance. $-$ indicates attribute dropped from the model due to multicollinearity.}
\end{tabular}%
}
\end{table*}

\begin{table*}
	\centering
    \caption{Results of logistic regression models non-newcomers. }
	\label{table:github_result-non-nc}
\begin{tabular}{|l|l|r|r|r|r|r|r|}
\hline
\multicolumn{2}{|c|}{\textbf{Dataset} $\rightarrow$}  & \multicolumn{2}{c|}{\textbf{GitHub(L)}}  & \multicolumn{2}{c|}{\textbf{GitHub(M)}}  & \multicolumn{2}{c|}{\textbf{GitHub(H)}} \\ \hline
\multicolumn{2}{|c|}{\textbf{Pseudo $R^{2} \rightarrow$}} & \multicolumn{2}{c|}{0.194$^{***}$} &  \multicolumn{2}{c|}{0.128$^{***}$} &  \multicolumn{2}{c|}{0.078$^{***}$} \\ \hline

 \textbf{Attribute} & \textbf{Reco(s)} & \textbf{OR$^{p}$} & \textbf{\textit{EP(\%)}}  &  \textbf{OR$^{p}$} & \textbf{\textit{EP(\%)}}  & \textbf{OR$^{p}$} & \textbf{\textit{EP(\%)}} \\ \hline
  log (code churn) & R1 & -- & -- & -- & -- & -- & --    \\ \hline                                
  GFI & R2 & \positive{  2.58$^{***}$ } & 0.01& \positive{  2.45$^{***}$ } & 0.01& \positive{  2.05$^{***}$ } & 0.01\\ \hline                  
  new feature & R3 & \negative{  $0.999^{***}$ } & 0.04& \negative{  $0.999^{***}$ } & 0.24& \negative{  $0.999^{***}$ } & 0.12      \\ \hline                                
  bug fix & R4 & \positive{  1.10$^{***}$ } & 0.01& \positive{  1.38$^{***}$ } & 0.29& \positive{  1.13$^{***}$ } & 0.05     \\ \hline  
  file count &R5 & -- & -- & -- & -- & -- & --    \\ \hline         
  entropy & R6 & \positive{  1.52$^{***}$ } & 0.30& \positive{  1.45$^{***}$ } & 0.26& \positive{  2.14$^{***}$ } & 0.88 \\ \hline 
  doc file & R7 &  \positive{  1.84$^{***}$ } & 0.23& \negative{  0.97$^{***}$ } & 0.01& \negative{  $0.999^{***}$ } & 0.01 \\ \hline
  log(insert) & R8& \negative{  0.86$^{***}$ } & 0.05& \negative{  0.92$^{***}$ } & 0.02& \negative{  0.85$^{***}$ } & 0.17 \\ \hline
  log(delete) & R9 & \positive{  1.14$^{***}$ } & 1.04& \positive{  1.08$^{***}$ } & 0.43& \positive{  1.11$^{***}$ } & 0.69  \\ \hline
  title length & R10 & \negative{  $0.999^{***}$ } & 0.01& \positive{  1.001$^{***}$ } & 0.05& \negative{  $0.999^{***}$ } & 0.17    \\ \hline
  desc. length & R11 & \negative{  $0.999^{***}$ } & 0.47& \negative{  $0.999^{***}$ } & 0.53& \negative{  $0.999^{***}$ } & 0.01 \\ \hline                                 
  title read. & R12 & \positive{  1.001$^{***}$ } & 0.01& \positive{  1.001$^{***}$ } & 0.01& \positive{  1.001$^{***}$ } & 0.01 \\ \hline                                  
 desc read. & R13 & \positive{  1.001$^{***}$ } & 0.01& \negative{  $0.999^{***}$ } & 0.01& \negative{  $0.999^{***}$ } & 0.69 \\ \hline                            
  promptness & R14 & \positive{  1.14$^{***}$ } & 3.83& \positive{  1.12$^{***}$ } & 2.71& \positive{  1.07$^{***}$ } & 0.68          \\ \hline  
  iteration & R15 &  \positive{  1.33$^{***}$ } & 3.25& \positive{1.19$^{***}$ } & 1.66& \positive{  1.06$^{***}$ } & 0.28         \\ \hline

\multicolumn{8}{p{11cm}}{ *** , **, and *  represent statistical significance at $p <$ 0.001, $p <$ 0.01, and $p <$ 0.05 respectively. No markers for $p$  indicate statistical insignificance.}
\end{tabular}
\end{table*}
\begin{table*}
	\centering
    \caption{Comparison of how each recommendation influences first patch acceptance for newcomers (NC) and non-newcomers (CT) for distinct projects. Symbols are interpreted as follows: ``\greencheck" $\rightarrow$ experimental results support the recommendation for the project, ``\contradict" $\rightarrow$ results contradict with the recommendation, ``\unsupported" $\rightarrow$ result is insignificant and ``--" $\rightarrow$ mapped variable was dropped or not present. \graybg{Gray} background indicates contradictory associations between newcomers and non-newcomers for that recommendation.}
    \vspace{-6pt}
	\label{table:recommendation_to_project}
	 \begin{tabular}{|l|c|c||c|c||c|c||c|c||c|c||c|c||c|c||c|c|}
\hline
\textbf{Project} $\rightarrow$ & \multicolumn{2}{c|}{\textbf{Android}}  & \multicolumn{2}{c|}{\textbf{LibreOffice}}  & \multicolumn{2}{c|}{\textbf{oVirt}} & \multicolumn{2}{c|}{\textbf{Typo3}} & \multicolumn{2}{c|}{\textbf{Wikimedia}} & \multicolumn{2}{c|}{\textbf{GitHub(L)}} & \multicolumn{2}{c|}{\textbf{GitHub(M)}} & \multicolumn{2}{c|}{\textbf{GitHub(H)}} \\ \hline

 \textbf{Reco. $\downarrow$ } & \textbf{NC}  & \textbf{CT} & \textbf{NC}  & \textbf{CT} & \textbf{NC}  & \textbf{CT} & \textbf{NC}  & \textbf{CT} & \textbf{NC}  & \textbf{CT} & \textbf{NC}  & \textbf{CT} & \textbf{NC}  & \textbf{CT} & \textbf{NC}  & \textbf{CT} \\ \hline \hline

 R2 & -- & -- & -- & -- & -- & -- & -- & -- & -- & -- & \unsupported & \greencheck & \greencheck & \greencheck & \greencheck & \greencheck 
\\ \hline
R3 & \greencheck & \greencheck & \unsupported & \unsupported & \unsupported & \unsupported & \unsupported & \unsupported & \unsupported & \unsupported & \unsupported & \contradict & \contradict & \contradict & \contradict & \contradict
\\ \hline
R4 & \greencheck & \greencheck & \unsupported & \unsupported & \greencheck & \greencheck & \greencheck & \greencheck & \greencheck & \greencheck & \greencheck & \greencheck & \greencheck & \greencheck & \greencheck & \greencheck
\\ \hline
R6* & \unsupported & \unsupported & \unsupported & \unsupported &  \cellcolor{gray!25} \greencheck & 
 \cellcolor{gray!25} \contradict & \unsupported & \unsupported & \unsupported & \unsupported & \cellcolor{gray!25} \greencheck & \cellcolor{gray!25} \contradict &  \cellcolor{gray!25}\greencheck &\cellcolor{gray!25}  \contradict & \cellcolor{gray!25} \greencheck & \cellcolor{gray!25} \contradict
\\ \hline
R7 & \greencheck & \greencheck & \cellcolor{gray!25} \greencheck & \cellcolor{gray!25} \contradict & \unsupported & \unsupported & \unsupported & \unsupported & \unsupported & \unsupported & \greencheck & \greencheck & \cellcolor{gray!25}\greencheck & \cellcolor{gray!25}\contradict & \cellcolor{gray!25} \greencheck & \cellcolor{gray!25} \contradict 
\\ \hline

R8* & \greencheck & \greencheck & \greencheck & \greencheck & \greencheck & \greencheck & \greencheck & \greencheck & \unsupported & \unsupported & \greencheck & \greencheck & \greencheck & \greencheck & \greencheck & \greencheck
\\ \hline

R9* & \contradict & \contradict & \unsupported & \unsupported & \contradict & \contradict & \unsupported & \unsupported & \unsupported & \unsupported & \cellcolor{gray!25} \greencheck & \cellcolor{gray!25} \contradict & \cellcolor{gray!25} \greencheck & \cellcolor{gray!25}  \contradict & \cellcolor{gray!25} \greencheck & \cellcolor{gray!25} \contradict
\\ \hline

R10 & \contradict & \contradict & \unsupported & \unsupported & \unsupported & \unsupported & \unsupported & \unsupported & \unsupported & \unsupported & \contradict & \contradict & \cellcolor{gray!25} \contradict & \cellcolor{gray!25} \greencheck & \contradict & \contradict 
\\ \hline
R11 & \contradict & \contradict & \unsupported & \unsupported & \unsupported & \unsupported & \contradict & \contradict & \unsupported & \unsupported & \contradict & \contradict & \contradict & \contradict & \cellcolor{gray!25} \greencheck & \cellcolor{gray!25} \contradict 
\\ \hline
R12 & \unsupported & \unsupported & \unsupported & \unsupported & -- & \unsupported & \unsupported & \unsupported & -- & \unsupported & \unsupported & \greencheck & \unsupported & \greencheck & \cellcolor{gray!25} \contradict & \cellcolor{gray!25} \greencheck
\\ \hline
R13 & \unsupported & \unsupported & -- & -- & \unsupported & \unsupported & \unsupported & \unsupported & \unsupported & \unsupported & \cellcolor{gray!25} \contradict & \cellcolor{gray!25} \greencheck & \contradict & \contradict & \contradict & \contradict
\\ \hline
R14 & \cellcolor{gray!25}  \contradict & \cellcolor{gray!25} \greencheck & \greencheck & \greencheck & \unsupported & \unsupported & \greencheck & \greencheck & \greencheck & \greencheck & \unsupported & \greencheck & \greencheck & \greencheck & \cellcolor{gray!25} \contradict & \cellcolor{gray!25} \greencheck
\\ \hline
R15  & \greencheck & \greencheck & \greencheck & \greencheck & \greencheck & \greencheck & \greencheck & \greencheck & \greencheck & \greencheck & \unsupported & \greencheck & \greencheck & \greencheck & \cellcolor{gray!25} \contradict &  \cellcolor{gray!25} \greencheck
\\ \hline
\multicolumn{17}{p{16cm}}{ * - interpretations for R6, R8, and R9 are opposite of the associations of code acceptance with mapped variables .}
 
\end{tabular}%
\end{table*}

\subsection{RQ3: Do OSS newcomers' onboarding recommendations also apply to non-newcomers?}
\label{sec:rq3}
Tables \ref{table:gerrit_result_nonnewcomer} and \ref{table:github_result-non-nc} show the results of our regression models for non-newcomers in Gerrit and GitHub projects. Recommendations (\textit{R4: bugfix}, \textit{R8: fewer line insertions}, and \textit{R14: proactive}) are supported across seven of eight datasets for non-newcomers. Recommendation \textit{R15: patient} is supported for all eight datasets. Recommendation \textit{R2: GFI} works for non-newcomers in GitHub projects. Recommendation \textit{R6: low complexity tasks} shows positive associations for three GitHub datasets and oVirt, which indicates with the increase in complexity, the patch acceptance probability increases for non-newcomers. \textit{R9: fewer line deletions} shows negative associations for non-newcomers, as with the increase in deleted lines, the patch acceptance probability increases. We also noticed a context-dependent nature for R3, R7, R10, and R13.  For example, \textit{R3: adding new features} positively correlates with Android acceptance but has negative associations among GitHub datasets. \textit{R7: adding documentation} increases the odds of acceptance in Android and GitHub (L) but lowers such odds in GitHub (M) and GitHub (H).

\vspace{-12pt}
\begin{boxedtext}
\textbf{Key takeaway 4:} \emph{Five recommendations-- R2: working on `good first issues', R4: Working on bug fixes, R8: working on changes requiring fewer line insertions,  R14: being proactive, and R15:  being patient, increases the odds of acceptance for non-newcomers\revision{'} changes.}
\end{boxedtext}

\vspace{4pt}
\noindent \textbf{Similarities and differences between recommendations: newcomers vs. non-newcomers}
Table \ref{table:recommendation_to_project} presents a side-by-side comparison of the recommendations for newcomers and non-newcomers across eight datasets. R1 and R5 are absent as their mapped variables were dropped for multicollinearity. 
Recommendations R4 and  R8 work for newcomers and non-newcomers in the Gerrit and GitHub ecosystems. Therefore, working on bug fixes and changes with fewer line insertions increases the odds of patch acceptance for both newcomers and non-newcomers.

Our results suggest that three recommendations (i.e., R6, R7, and R9) are newcomer-specific and do not apply to non-newcomers. Among the GitHub projects and OVirt, \textit{working on low complexity tasks (R6)} improves the odds of acceptance for newcomers, but opposite associations are seen for non-newcomers. Similarly, \textit{working on documentation (R7)} increases the odds of acceptance for newcomers to LibreOffice, GitHub (M), and GitHub (L) but decreases the odds for non-newcomers from those contexts. Another contradiction was observed among GitHub projects for \textit{R9: fewer line deletions} since it works for newcomers but not for non-newcomers. 

We also noticed context-dependence for \textit{R14: being proactive} and \textit{R15: being patient}. While R14 works for non-newcomers to Android and GitHub (H), it shows the opposite for non-newcomers in those contexts. Similarly, a higher number of iterations increases the odds of acceptance for non-newcomers to GitHub (H) but decreases such odds for newcomers. 

Surprisingly, four recommendations related to longer and readable titles/ descriptions (i.e., R10, R11, R12, and R13) are either contradicted or insignificant in most cases. However, our results indicate contradictions between newcomers and non-newcomers of GitHub for these four recommendations. \textit{R10: writing longer titles} increases the odds of acceptance for non-newcomers of GitHub (M) but decreases for newcomers. \textit{R11: Writing longer descriptions} increases the odds of acceptance for newcomers to GitHub (H), but it's the opposite for non-newcomers. \textit{R12: writing a readable title} increases the odds of acceptance for non-newcomers of GitHub (H) but decreases for newcomers. Finally, \textit{R13: writing readable description} applies to non-newcomers of GitHub (L) but is the opposite for newcomers.

\vspace{-12pt}
\begin{boxedtext}
\textbf{Key takeaway 5:} \emph{Two recommendations-- R4: Working on bug fixes and R8: working on changes requiring fewer line insertions apply to newcomers and non-newcomers. Three recommendations -- R6: working on low complexity changes, R7: working on documentation, and R9: working on changes requiring fewer line deletions, are mostly newcomer-specific and show contradictory associations between newcomers and non-newcomers. Our results also found context-dependent differences for six recommendations. }
\end{boxedtext}

\section{Discussion and Implications}
\label{sec:discussion}
The following subsections discuss our findings and suggest key implications.

\subsection{Observed differences between Gerrit and GitHub-based projects}
\label{gerrit-github-diff}
Our results suggest a few differences between Gerrit and GitHub-based projects. However, these few differences may be primarily due to project characteristics rather than platforms. 
The most noticeable difference is  \textit{R9:  working on changes requiring fewer lines to be deleted}. While this recommendation applies to newcomers to GitHub projects, it shows contradictory associations for Android and oVirt and is insignificant for the remaining three Gerrit-based projects. These results indicate that deleting multiple lines does not significantly reduce the odds of newcomers' patch acceptance for matured OSS projects, such as the ones using Gerrit. However, these odds decrease on GitHub.

\vspace{4pt}
\subsection{Comparison with Tan \textit{et al.}}
Five out of the 15 factors investigated by this study were also included in Tan \textit{et al.}'s\cite{tan2020first} investigation of GFIs on GitHub.  
Similar to their findings, our results suggest that the odds of acceptance significantly decrease with the \textit{number of lines deleted (R9)}. However, although they found no significant correlation between effective GFIs and \textit{the number of lines added (R8)}, our results suggest a significant positive association. Moreover, their results suggest \textit{description length (R11)} increases the likelihood of a GFI being effective. Our results found a context dependence with GitHub (H) supporting a positive association between description length and patch acceptance, but GitHub (L) and GitHub (M) show opposite associations.
Finally, Tan \textit{et al.}'s\cite{tan2020first} also found no significant association between GFI effectiveness and \textit{title length (R10)} / \textit{description readability (R13)}. However, our results found those attributes negatively associated with newcomers' PR acceptance on GitHub. However, these observed differences may be due to the differences in samples and analysis goals, as they aim to identify differences between GFIs completed by newcomers and those completed by non-newcomers. However, we differentiate between accepted and rejected PRs submitted by newcomers, including GFIs and non-GFIs.

\subsection{Implications}
\label{sec:implication}

\textbf{Recommendations, applicable for both Newcomers and Non-newcomers:} Our results indicate three \textit{task selection} recommendations apply to both newcomers and non-newcomers, which are \textit{working on good first issues (R2)}, \textit{bug fixes (R4)},  and \textit{changes requiring fewer line insertions (R8)}.  Since the odds of a newcomer's code acceptance significantly decrease with the number of lines added, we suggest OSS mentors split big tasks into smaller ones, and if the task is intended for newcomers \cite{subramanian2020analyzing}. 
\textit{GFI (R2)} tags significantly improve newcomers' odds of accepting a patch. For example, among the GitHub (H) group projects, a newcomer working on GFI has 5.57 times higher odds of accepting change than working on non-GFIs. Therefore, we recommend that newcomers start with GFIs and that project maintainers use GFI-related tags to attract and successfully onboard newcomers. 
Our results also indicate that newcomers' bug fixes have significantly higher odds of acceptance than other changes.  For example, newcomers' bug fixes in Android have 2.41 times higher odds of getting accepted than other types of changes (Table \ref{table:gerrit_result_newcomer}). 
Since these four recommendations also apply to non-newcomers, a contributor can continue following these recommendations for long-term success with an OSS project.

\vspace{2pt} \noindent
\textbf{Newcomer-specific recommendations:}
Our results suggest that the odds of a non-newcomer's change getting accepted increase with its complexity across GitHub and oVirt, but opposite associations are seen for a newcomer's change. Hence, once newcomers get a few patches accepted, we recommend gradually taking more complex ones to increase their odds of becoming long-term contributors. 
Similarly, while newcomers to LibreOffice, GitHub(M), and GitHub (H) have higher odds of acceptance with \textit{documentation-related changes (R7)}, the opposite odds are found for non-newcomers to those projects.
 Therefore, after a few documentation changes are accepted, we recommend that joiners of those projects focus on non-documentation changes if they want to become long-term members. Finally, while the odds of code acceptance decrease with the \textit{number of line deletions (R9)} for newcomers on GitHub, the opposite is true for non-newcomers on GitHub. Hence, we recommend that newcomers to GitHub-based OSS projects start with changes that require fewer line deletions. However, prospective joiners do not need to comply with this restriction after getting a few of their PRs accepted.

\vspace{2pt} \noindent
\textbf{Insignificant Recommendations :} \textit{R3: adding a new feature} increases the odds of newcomers' patch acceptance only in Android. For the remaining four Gerrit-based projects, this recommendation is inconclusive. Moreover, we found that this recommendation decreases the odds of acceptance of newcomers' PR on GitHub. Hence, we recommend that newcomers, particularly those contributing to OSS projects on GitHub, do not start with new features. Our results suggest that \textit{title length (R10)}, \textit{title readability (R12)}, and \textit{description readability (R13)} are either insignificant or negatively associated with newcomers' code acceptance.  Hence, we recommend that newcomers focus on code quality rather than spending a lot of effort to improve the title /description's readability.

\vspace{2pt} \noindent
\textbf{ Context-dependent recommendations:} 
Two \textit{task completion} recommendations, \textit{R14: being proactive} and \textit{R15: being patient}, indicate positive associations with code acceptance for both newcomers and non-newcomers in almost all contexts, with few exceptions.
\textit{R15: Being patient} by following up with iterations to address reviewers' suggestions is contradicted for newcomers to GitHub (H) but supported across all other cases. We further investigated this result since we expected this recommendation (R15) to be universal. 
OSS projects belonging to GiHub (H) have high development activity with more than 32 PRs per month. Therefore, maintainers of these projects may not have adequate time to go back and forth to suggest improvements to make a change acceptable. If a change is acceptable within a few iterations, they will accept it; otherwise, it will get rejected. For the same reason, \textit{R14: being proactive} shows a contradictory association for GitHub (H), as maintainers of highly active projects look for correct implementations, not quick trial and error efforts.
Based on these results, we recommend that newcomers and non-newcomers be proactive and patient and promptly address suggestions from reviewers. However, when submitting first changes to highly active projects such as the ones from GitHub (H) or Android, newcomers should focus on getting the change correct within a few iterations, as maintainers of those projects may not have time to do multiple rounds of reviews.

Our results suggest that the odds of acceptance for newcomers' PRs increase with \textit{description length (R11)} among the OSS projects in GitHub (H). However, such an association is not present in any other context. GitHub (H) includes popular and highly active OSS projects. Since the maintainers of those projects are busy, detailed descriptions for PRs help their reviews and increase the odds of acceptance. Hence, we recommend that newcomers add detailed descriptions, especially when opening PRs for highly active projects. 

\vspace{2pt} \noindent
\textbf{Future directions:} The results of our study found three newcomer-specific recommendations related to task selection,  suggesting that an OSS contributor should change his/her task selection strategies after his/her first patches are accepted in an OSS project if he/she intends to become a long-term contributor to that project. 
However, non-newcomers groups are diverse, as prior  OSS literature has further categorized non-newcomers into various groups such as quasi-contributors~\cite{steinmacher2018almost}, casual contributors~\cite{pinto2016more},  episodic volunteers~\cite{barcomb2020managing}, core and peripheral developers. Based on our results, it is plausible that task-related strategies may further differ among various non-newcomer groups. Such investigations will allow customizing recommendations for a non-newcomer based on his/her individual goals. We consider those analyses as one of the future directions for this research.

\section{Threats to Validity}
\label{sec:validity}
\vspace{4pt} \noindent \textbf{Internal Validity}
The primary threat to internal validity is the identification of primary studies for our literature review (SLR). Although we have followed the recommended guidelines to conduct the SLR and validated our lists based on multiple experts' recommendations, we may have missed a few papers. Although we do not believe many papers are missing from our list, this threat remains. Regardless, our identification of 15 recommendations and validation of those using large-scale datasets provides valuable insights.

\vspace{4pt} \noindent \textbf{External Validity}
We collected code review comments from 5 popular OSS projects and mined 1.4 million code review requests from 5 Gerrit projects and 5.8 million pull requests from 1,155 GitHub projects.  For Gerrit projects, we ensured each of our selected projects has at least 50,000 completed CRs. From GitHub, we selected projects that use a popular programming language, have 20 contributors, and have at least 20 pull requests in the last three months.
 Although we have selected a large number of projects using two different categories of onboarding platforms, our sample may not be representative of the entire OSS spectrum. We have made our dataset and scripts publicly available to promote replication of our results in other settings.

\vspace{4pt} \noindent \textbf{Construct Validity}
Our selection of quantitative metrics to assess each recommendation is the primary threat to construct validity. While some recommendations are straightforward and can be directly mapped to a measure (i.e., longer titles or the number of lines added), recommendations such as complexity and proactivity are subject to interpretation. A reader may disagree with our selection and consider another metric a more accurate recommendation measure. Regardless, we believe that our metrics represent reasonable measures, and this threat is minimal since most of those have been used in prior code review-related studies to measure similar constructs.

\vspace{4pt} \noindent \textbf{Conclusion Validity}
For some regression modeling, the normality assumption affects a model's performance. According to the normality assumption, the dependent variable must be distributed normally. If not, a mathematical transformation is required to transform the dependent variable into a normal distribution. To avoid such a scenario, we have constructed our MLR model using the Generalized Linear Model (GLM) with \code{"binomial"} family. Since GLM is resilient to normality assumptions, this threat does not apply to our analyses \cite{lee1996hierarchical}. Moreover, we used mature libraries for modeling and followed the recommended guidelines~\cite{harrell2015regression}. Therefore, we do not anticipate any significant threat to the conclusion's validity.

Combining various categories of non-newcomers into one group to train one regression model poses a threat to conclusion validity, since non-newcomers are diverse~\cite{lee1996hierarchical,barcomb2020managing,crowston2003defining,pinto2016more}. Hence, some recommendations may apply differently to a non-newcomer subgroup than the whole. A fine-grained investigation is essential to identify those nuanced differences.

\section{Conclusion}
\label{sec:conclusion}

In this two-stage mixed-method study, we first conducted an SLR. We identified 15 task-related actionable recommendations that newcomers to OSS projects can follow to improve their odds of successful onboarding. In the second stage, we conduct a large-scale empirical study to assess whether those recommendations assist newcomers' successful onboarding. Our results suggest that four recommendations positively correlate with newcomers' first patch acceptance in most contexts. Four recommendations are context-dependent, and four indicate significant negative associations for most projects. Our results also found three newcomer-specific recommendations: working on -- i) low complexity changes, ii) documentation, or iii)  changes requiring fewer line deletions. 
Therefore, a contributor should abandon these three strategies at non-newcomer status to increase the odds of successful onboarding. 

\section*{Acknowledgment} 

Work conducted for this research is partially supported by the US National Science Foundation under Grants 1850475 and 2340389. Any opinions, findings, conclusions, or recommendations expressed in this material are those of the author(s) and do not necessarily reflect the views of the National Science Foundation.

\balance

\bibliographystyle{IEEETran}
\balance
\bibliography{references}

\end{document}